\documentclass[final,5p,times]{elsarticle}

\usepackage{amsmath,amsthm}
\usepackage{amssymb,latexsym}
\usepackage{graphicx}
\usepackage{setspace}
\usepackage[mathscr]{eucal}
\usepackage{subfig}

\numberwithin{equation}{section}

\journal{$\qquad$ $\qquad$ $\qquad$ $\qquad$ $\qquad$ $\qquad$ Physics Letters A 377, 1434-1438 (2013)}

\begin{document}

\begin{frontmatter}


 \author{Stefan C. Mancas \fnref{label1}}
   \author{Haret C. Rosu\fnref{label2}}

\title{Integrable Dissipative Nonlinear Second Order Differential\\ Equations via Factorizations and Abel Equations}

 \address[label1]{Department of Mathematics, Embry--Riddle Aeronautical University, Daytona Beach, FL. 32114-3900, U.S.A.} 
 \address[label2]{IPICYT, Instituto Potosino de Investigacion Cientifica y Tecnologica,\\Apdo Postal 3-74 Tangamanga, 78231 San Luis Potos\'{\i}, S.L.P., Mexico.}

\begin{abstract}
\it{We emphasize two connections, one well known and another less known, between the dissipative nonlinear second order differential equations and the Abel equations which in its first kind form have only cubic and quadratic terms. Then, employing an old integrability criterion due to Chiellini, we introduce the corresponding integrable dissipative equations. For illustration, we present the cases of some integrable dissipative Fisher, nonlinear pendulum, and Burgers-Huxley type equations which are obtained in this way and can be of interest in applications. We also show how to obtain Abel solutions directly from the factorization of second-order nonlinear equations.}
\end{abstract}


\end{frontmatter}

\section{Introduction}
The connections between second order differential equations of linear and nonlinear type and Riccati and Abel first order nonlinear equations, respectively, are well known and referred in many textbooks \cite{Davis,BO}. Abel equations of the first kind look similar to the Riccati equations but with an additional cubic nonlinearity,
\begin{equation}\label{fkind}
\frac{d y}{du}
=\varphi_0(u)+\varphi_1(u)y+\varphi_2(u)y^2+\varphi_3(u)y^3~,
\end{equation}
while those of the second kind have the more general format,
\begin{equation}\label{n2p1}
[g_0(u)+g_1(u)\eta]\frac{d \eta}{du}
=f_0(u)+f_1(u)\eta+f_2(u)\eta^2+f_3(u)\eta^3~,
\end{equation}
and also chronologically occurred  first in Abel's famous study on elliptic functions \cite{Abel} in the form:
\begin{equation}\label{ab1829}
[g_0(u)+\eta]\frac{d \eta}{du}
=f_0(u)+f_1(u)\eta+f_2(u)\eta^2~.
\end{equation}
 However, we stress that the two types of Abel equations are equivalent since they can be obtained one from the other through the change of the dependent variable $g_0(u)+g_1(u)\eta=1/y$. In the following, we first state a simple `connection' result for the particular case of (\ref{n2p1}) as given in (\ref{n2p}). We then turn the latter in its first-kind form, equation (\ref{n19}) below, and present Chiellini's integrability result \cite{chie31} which is also mentioned in the book of Kamke \cite{kam}. Our discussion of Chiellini's result goes beyond that in Chiellini's paper and in Kamke's book. Next, we show that one does not need to find $\eta$ from (\ref{n2p}), but directly from a factorization technique as applied to the corresponding second order ODE. Finally, we will use our findings to introduce new types of integrable second-order nonlinear ODE providing also their solutions and we end up the Letter with several conclusions. For the reader interested in other approaches to the integrability of Abel's equation and general overview we recommend \cite{clr}.

\section{Two fundamental results}

We now briefly present the results we need.
A first fundamental result is the following lemma.

\begin{quote}
{\bf Lemma 1}: Solutions to a nonlinear second order ODE of the type
\begin{equation}\label{n1p}
\frac{d^2u}{d \zeta^2}+g(u) \frac{du}{d \zeta}  +h(u)=0,
\end{equation}
where $g$ and $h$ are general functions of $u(\zeta)$, may be obtained via the solutions to the second-kind Abel's equation
\begin{equation}\label{n2p}
\eta\frac{d \eta}{du}+g(u)\eta+h(u)=0~,
\end{equation}
 and vice versa using the following relationship
\begin{equation}\label{n10}
\frac{du}{d\zeta}=\eta(u(\zeta))~.
\end{equation}
\end{quote}
This equivalence can be found in the book of Polyanin and Zaitsev in the simpler case $h(u)=u$, which they call nonlinear oscillator equations \cite{pz}.

{\bf Proof}: To show the equivalence, one just need the simple chain rule
\begin{equation}
\frac{d^2u}{d\zeta^2}=\frac{d \eta}{d u}\frac{du}{d\zeta}
\end{equation}
which turns (\ref{n1p}) into
\begin{equation}
\frac{d \eta}{du}\frac{du}{d \zeta}+g(u) \frac{du}{d \zeta}+h(u)=0,
\end{equation}

 which is (\ref{n2p}).

\medskip
Thus, the solutions of the several classes of known solvable Abel equations \cite{ctr1} are just the derivatives of the solutions of the second order nonlinear equations of the type (\ref{n1p}). Hence, if one could solve (\ref{n2p}) for $\eta$, then $u$ could be found via inverting
\begin{equation}
\int^u \frac{1}{\eta}dr=\zeta - \zeta_0~.
\end{equation}

\medskip

As we already said we will also need the following result.
\begin{quote}
{\bf Chiellini's integrability condition:}
 If $g(u)$ and $h(u)$ are connected by
\begin{equation}\label{n16}
\frac{d}{du}\left(\frac{h(u)}{g(u)}\right)=kg(u),
\end{equation}
for some constant $k$, then Abel's equation is integrable.
\end{quote}

{\bf Proof}: Letting $\eta(u)=\frac{1}{y(u)}$, then equation~(\ref{n2p}) becomes
\begin{equation}\label{n19}
\frac{dy}{du}=h(u)y^3+g(u)y^2~,
\end{equation}
which is an Abel equation of the first kind without linear and free terms. We wish to point out that an Abel equation of the type (\ref{n2p}) with an additional quadratic (Riccati) nonlinearity is still allowed in the connection result because it leads to a linear term $l(u) y$ in the Abel equation \eqref{n19} and such a term can be always eliminated via the transformation $y=e^{\int l(u)du}\hat y$. Hence without loss of generality, we will work only with quadratic and cubic nonlinearities.

Furthermore, let $z=y\frac{h}{g}$ and take into account Chiellini's condition, then one gets
\begin{equation}
\frac{dz}{du}=\frac{h}{g}\frac{dy}{du}+kgy
\end{equation}
 and, therefore, the first-kind Abel equation (\ref{n2p}) becomes
 \begin{equation}\label{n20}
\frac{dz}{du}=\frac{g^2}{h}\left(z^3+z^2+kz\right)
\end{equation}
which is separable as follows
\begin{equation}\label{n21}
\int \frac{dz}{z(z^2+z+k)}=\int \frac{g^2}{h}du~.
\end{equation}
The right-hand-side of (\ref{n21}) can be written as $\frac{1}{k}\int \frac{d(\frac{h}{g})}{\frac{h}{g}}$.

Therefore, one obtains
\begin{equation}\label{n22}
\int \frac{dz}{z(z^2+z+k)}=\frac{1}{k}\ln \big| \frac{h}{g}\big |+c~.
\end{equation}
The implicit solutions depend on the constant $k$, and are given by
\begin{equation}
\left\{ \begin{array}{ll}
|z|^k\frac{|z-z_1|^{z_2}}{|z-z_2|^{z_1}}=d_0\big|\frac h g\big|^{\sqrt{1-4k}}  & \textrm{if $k<\frac 1 4 $}\\
e^{\left[\frac {1}{1+2z}-2 \arctan (1+4z)\right]}=d_1\frac h g & \textrm{if $k=\frac 1 4 $}\\
\ln \frac{|z|}{\sqrt{z^2+z+k}}-\frac{1}{\sqrt{4k-1}}\arctan \frac{2z+1}{2\sqrt{4k-1}}=\ln d_2\big|\frac h g\big|& \textrm{if $k>\frac 1 4 $}~.
\end{array} \right.
\end{equation}
$d_0$, $d_1$, $d_2$ are general integration constants, while $z_1$ and $z_2$ are the distinct real roots of $z^2+z+k=0$ for $k<\frac 1 4$.
\medskip

Along the years, this important result drew the attention of very few people. In the 1960s, Bandi\'c wrote a couple of related mathematical papers \cite{Bandic1,Bandic2}, while much later Borghero and Melis \cite{bm90} used it in the so-called Szebehely's problem of finding the generalized potential function which is compatible with prescribed dynamical trajectories of a holonomic system. More recently, Mak and Harko \cite{mh02,mh2,mh3} devised a method to get general solutions of the first-kind Abel equation from a particular solution based on Chiellini's result and the list closes with a paper by Yurov and Yurov in cosmology \cite{YY}.
We also mention a very recent unpublished work by Harko et al \cite{hlm} for the case of particular Li\'enard equations.

\section{A factorization method}

In this section, we will explain how we use the factorization method applied  to (\ref{n1p}) to obtain the solutions of (\ref{n2p}).

The factored form of (\ref{n1p}) reads
\begin{equation}\label{n2}
\left [\frac{d}{d \zeta}-\phi_{2}(u)\right] \left[\frac{d}{d \zeta}-\phi_{1}(u)\right ]u(\zeta)=0~.
\end{equation}
Expanding (\ref{n2}) and identifying terms, Rosu and Cornejo-P\'erez \cite{rosu1,rosu2} obtained the equation
\begin{equation}\label{n4}
\frac{d^2u}{d\zeta^2}-\left(\phi_{1}+\phi_{2}+
\frac{d\phi_{1}}{du} u \right)\frac{du}{d\zeta}+\phi_{1}\phi_{2}u=0~,
\end{equation}
which leads to the following conditions on the two factoring functions
\begin{eqnarray}\label{n5}
&&\phi_{1}\, \phi_{2}=\frac {h(u)}{u},\\
&&\phi_{1}+\phi_{2}=- \frac{d\phi_{1}}{du} \, u -g(u)~.
\end{eqnarray}

The solutions $\phi_1(u)$ and $\phi_2(u)$ of the system (\ref{n5}) are easily obtained by solving the quadratic equation
\begin{equation}\label{n6}
t^2-St+P=0,
\end{equation}
 where $S= -\frac{d\phi_{1}}{du} \, u -g(u)$, and $P=\frac {h(u)}{u}$, and $t^{\pm} =\phi_{1,2}$.
 By choosing $t^{+}=\phi_1$, we obtain
\begin{equation}\label{n7}
\phi_1\left(\phi_1+\frac{d \phi_1}{du}u \right)+g\phi_1+\frac{h}{u}=0.
\end{equation}
It is not a coincidence to notice that if
\begin{equation}\label{n6p}
\eta(u)=u\phi_1(u)~,
\end{equation}
then the equation for factors (\ref{n7}), is indeed Abel's equation (\ref{n2p}).
Therefore, rather than solving (\ref{n2p}), $\eta$ can be obtained from the factors of the ODE.
Interestingly, the factorization method provides another argument for the two equivalences, (\ref{n10}) and (\ref{n6p}). From the factorization (\ref{n2}) we have $\left [\frac{du}{d \zeta}-\phi_{1}(u)\right]u(\zeta)=0$, i.e., $\frac{du}{d \zeta}=\phi_{1}(u)u$. Thus, if we take $ \frac{du}{d \zeta}=\eta$, then also $\phi_{1}(u)u=\eta$.
In addition, the interpretation of the Abel solution as in (\ref{n6p}) 
permits the formulation of another lemma as follows.

\begin{quote}
{\bf Lemma 2}: For Chiellini-integrable ODEs, i.e., ODEs that have $g(u)$, and $h(u)$ connected via Chiellini's condition, the solution to Abel's equation (\ref{n2p}) is given by
\begin{equation}\label{sol}
\eta(u)=c_k\frac{h(u)}{g(u)}~,
\end{equation}
where the constant $c_k$ is given in terms of Chiellini's constant through
\begin{equation}\label{sol-1}
c_k=\frac{-1\pm\sqrt{1-4k}}{2k}~.
\end{equation}
%

{\bf Proof}: Let $\eta(u)=c_k\frac{h(u)}{g(u)}$, and substitute in (\ref{n2p}), then one gets $kc_k^2+c_k+1=0$ with the roots given by (\ref{sol-1}).
\end{quote}

Lemma 2 is very useful because one can employ it to find $\eta$ from $g(u)$ and $h(u)$ as follows.
\begin{quote}

{\bf Theorem:}
For an integrable ODE of type (\ref{n1p}),
\begin{enumerate}
\item[i)]  if $g(u)$ is known, then
 \begin{equation}\label{n17}
\eta_{g}(u)=c_k\left(c_0+k \int^u g(r)dr\right)
\end{equation}
or
 \item[ii)] if $h(u)$ is known, then
\begin{equation}\label{n18}
\eta_{h}(u)=\pm c_k\sqrt{c_1+2k \int^u h(r)dr}
\end{equation}
\end{enumerate}

{\bf Proof}: For both cases we will use Chiellini's condition together with lemma 2. For simplicity, we put $c_k=1$.
\begin{enumerate}
\item[i)]  if we have $g(u)$, then
 \begin{equation}\label{n30}
h(u)=g(u)\left(c_0+k \int^u g(r)dr\right)
\end{equation}
by integrating (\ref{n16}).
 \item[ii)] If we know $h(u)$, then we multiply equation~(\ref{n16}) by $\frac{h}{g}$ to get
\begin{equation}\label{n31}
\frac{h}{g} \frac{d}{du}\left(\frac{h}{g}\right)=kh.
\end{equation}
By integrating once with respect to $u$, we obtain

\begin{equation}\label{n32}
\frac{h^2}{g^2} =c_1+2k  \int^u h(r)dr.
\end{equation}
\end{enumerate}

\end{quote}

\medskip

\section{Examples of integrable dissipative equations obtained from the above Theorem}

We use now these results to obtain four integrable dissipative equations of type (\ref{n1p}), either by starting with given $g$ or $h$.
We prefer to begin with the two cases of given $h$ because usually the nonlinear equations are identified by their nonlinear term(s) and not so much by their dissipation coefficient. In these illustrative examples, we will take $c_k=1$, which corresponds to $k=-2$.

\medskip
\subsection{Dissipative Fisher's equation}
\noindent In this case, let  $h(u)=u(1-u)$. Using the theorem,
\begin{equation} \label{f-2}
\eta_h(u)=\sqrt{c_1-2u^2+\frac{4u^3}{3}}~.
\end{equation}

Then, one gets the following integrable dissipative Fisher's equation
\begin{equation}\label{f-3}
u_{\zeta \zeta}+\frac{u(1-u)}{\sqrt{c_1-2u^2+\frac{4u^3}{3}}}u_{\zeta}+u(1-u)=0~,
\end{equation}
with closed form solution given by
\begin{equation}\label{f-3p}
\zeta-\zeta_0=\frac{\sqrt 3}{2}\int^u \frac{dr}{\sqrt{r^3-\frac{3}{2}r^2+c_2}}~,
\end{equation} 
where $c_2=\frac{3}{4}c_1$.

We notice that (\ref{f-3}) can be also written in the convective form
\begin{equation}\label{f-4}
u_{\zeta \zeta}+\mu(u)u u_{\zeta}+u(1-u)=0~,
\end{equation}
where $\mu(u)=\frac{1-u}{\sqrt{c_1-2u^2+\frac{4u^3}{3}}}$ can be interpreted as a tuning function of the convection.

In general, convective Fisher equations have been applied with interesting results in population dynamics \cite{gk03}, while the case
of constant $\mu$ has been studied by Sch\"onborn and collaborators \cite{s94}.

The dissipative Fisher solution of (\ref{f-3}) for $c_2=\frac{1}{2}$ is displayed in Fig.~ 1, and it has the dark soliton profile $u(\zeta)=1-\frac{3}{2}{\rm sech}^2 \frac{\zeta}{2}$. On the other hand, when $c_2=\frac{1}{4}$ the solution is the Jacobi elliptic sn function of the form $u(\zeta)=\frac{1-\sqrt{3}}{2}+\sqrt{3}\,{\rm sn}^2\left(\frac{\zeta}{3^{1/4}}|2\right)$
as plotted in Fig. ~2.

\medskip

\subsection{Dissipative nonlinear pendulum equation}
\noindent For this case, $h(u)=\sin u$ and then $\eta_h(u)$ is given by
\begin{equation}\label{sg-2}
\eta_h(u)=\sqrt{c_3+4 \cos u}~.
\end{equation}
Therefore, one gets the dissipative integrable nonlinear pendulum equation
\begin{equation}\label{sg-3}
u_{\zeta \zeta}+\frac{\sin u}{\sqrt{c_3+4 \cos u}}u_{\zeta}+\sin u=0~,
\end{equation}
with closed form solution in terms of the elliptic integral of the first kind $F(u|m)$
\begin{equation}\label{sg-4}
\zeta-\zeta_0= \frac{\sqrt {2m}}{2}\int ^u\frac{d \theta }{\sqrt{1-m\sin^2 \theta}}=\frac{\sqrt {2m}}{2}F(u|m)~,
\end{equation}
where $m=\frac{8}{c_3+4}$.

\medskip

In the case of the dissipative nonlinear pendulum equation (\ref{sg-3}) for $m=2$ the solution is the amplitude for Jacobi elliptic function $u(\zeta)={\rm am}(\zeta |2)$, where $\zeta=F(u|2)$ as displayed in Fig.~3.  If $m=1$, then $F(u|1)=\ln|\sec(u)+\tan(u)|$, and hence the solution to equation (\ref{sg-3}) is \\ $u(\zeta)=\pm {\rm arc sin}\big(\tanh(\sqrt {2}\zeta)\big)$, see Fig.~4. If $m=8/9$, then $u(\zeta)={\rm am}(\frac{3 \zeta}{2}|\frac{8}{9})$, see Fig.~5.

\medskip

\subsection{Generalized nonlinear pendulum equation with sine dissipation}

\noindent If we take $g(u)=\sin u$, then (\ref{n30}) gives $h(u)=c_0\sin u +\sin 2u$, and therefore this is a generalized nonlinear pendulum equation of the type
\begin{equation}\label{sine-diss}
u_{\zeta\zeta}+\sin u u_\zeta +c_0\sin u +\sin 2u=0.
\end{equation} The solutions are obtained by inverting
\begin{equation}\label{last}
\zeta=\int \frac{du}{c_0+2\cos u}~,
\end{equation}
which leads to the following solutions
$$
u(\zeta) = \left\{ \begin{array}{ll}
2 \arctan {(2 \zeta)} & \textrm{if $c_0=2$}\\
2 \rm{arccotan} {(2 \zeta)} & \textrm{if $c_0=-2$}\\
2 \arctan{\Big(\frac{(2+c_0)\tanh{(\frac1 2 \sqrt{4-c_0^2}\zeta})}{\sqrt{4-c_0^2}}}\Big)  & \textrm{if $|c_0|<2$}\\
2 \arctan{\Big(\frac{(2+c_0)\tan{(\frac1 2 \sqrt{4-c_0^2}\zeta})}{\sqrt{4-c_0^2}}}\Big) & \textrm{if $|c_0|>2$}~.
\end{array} \right.
$$

Plots of the last two cases for $c_0=1$ and $c_0=3$ are displayed in Figs.~6 and 7, respectively.

%
\subsection{Burgers-Huxley type equation}

\noindent An equation of this type can be obtained if we let $g(u)=\mu u$, which through (\ref{n30}) leads to
\begin{equation}\label{bhnl}
h(u)=\mu^2 u(\sqrt{c_0/\mu}- u)(\sqrt{c_0/\mu}+u)~.
\end{equation}
 This is similar to a Huxley nonlinearity, although not for the typical range of the Huxley parameters. Such equations reflect the complex interplay between the nonlinearity, convection, and diffusive transport for waves propagating in biological and chemical systems \cite{zhou,wang}. The solutions are obtained from
\begin{equation}\label{pen}
\zeta=-\frac{1}{\mu}\int^u \frac{dr}{r^2-\frac{c_0}{\mu}}~.
\end{equation}
After inverting, this leads to three simple elementary  solutions as follows

$$
u(\zeta) = \left\{ \begin{array}{ll}
\left(\frac{c_0}{\mu}\right)^{1/2}\tanh (\sqrt{\mu c_0 }\zeta) & \textrm{if $\frac{c_0}{\mu}>0$}\\
\,\,\,\frac{1}{\mu \zeta} & \textrm{if $c_0=0$}\\
\left(-\frac{c_0}{\mu}\right)^{1/2}\tan (\sqrt{-\mu c_0 }\zeta) & \textrm{if $\frac{c_0}{\mu}<0$}~.
\end{array} \right.
$$

We do not plot these solutions because they are well-known elementary functions.

One may think about a medium, e.g., a biological membrane, with convection tuned by the $\mu$ parameter and with the symmetric Burgers-Huxley nonlinearity implied by the Chiellini integrability as given in (\ref{bhnl}). Then the amplitude of the switching solution (for positive $c_0/\mu$) is inverse proportional to $\sqrt{\mu}$. Thus, stronger convection leads to a less pronounced switching effect in this case.

\bigskip

\section{Conclusion}
 In summary, we have shown that the connections between dissipative nonlinear second order differential equations and the integrable Abel equations can be very useful to extend the class of integrable dissipative nonlinear equations. The convective-like Fisher's equation and the dissipative nonlinear pendulum equation as well as the Burgers-Huxley type equation introduced here are such examples but many other equations can be generated in this way. All these equations may be thought as designed ones, i.e., they have either special dissipation factors for given nonlinearities or special nonlinearities for given dissipation such that they have well defined solutions.
In addition, we showed how one can get Abel solutions directly from the factorization of the second-order nonlinear equations.

\bigskip

\noindent Acknowledgment: The second author thanks CONACyT-Mexico for a sabbatical fellowship.

\begin{figure}[x]
  \centering
  \includegraphics[width= 6.08 cm, height=6.07 cm]{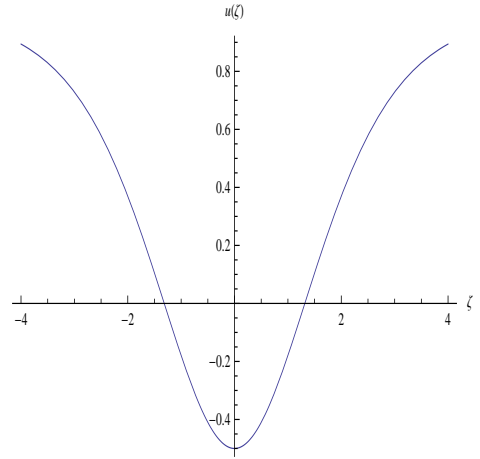} 
  \caption{(Color online) Solution with dark soliton profile, $u(\zeta)=1-\frac32{\rm sech}^2{\frac{\zeta}{2}}$, of the Abel-dissipative Fisher equation (\ref{f-3}) for $c_1=\frac23$.}
  \label{S-fig1}
 \end{figure}
 \vfill
\begin{figure}[x]
  \centering
  \includegraphics[width= 6.08 cm, height=6.07 cm]{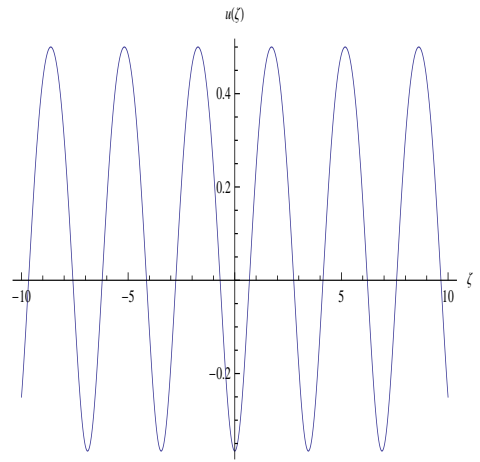}
  \caption{(Color online) Elliptic function solution $u(\zeta)=\frac{1-\sqrt{3}}{2}+\sqrt{3}{\rm sn}^2\left(\frac{\zeta}{3^{1/4}}|2\right)$ of the Abel-dissipative Fisher equation (\ref{f-3}) for $c_1=\frac13$.}
  \label{S-fig2}
 \end{figure}

 \vfill

\begin{figure}[x]
  \centering
  \includegraphics[width= 6.08 cm, height=6.07 cm]{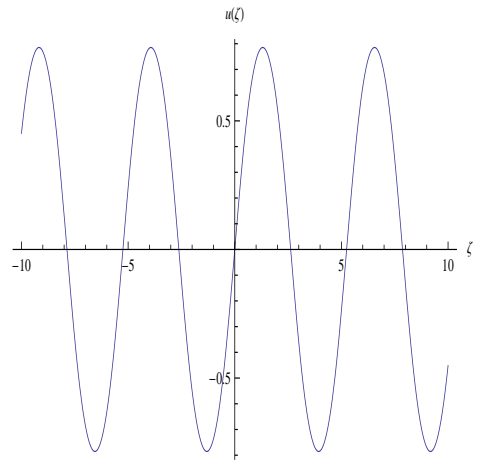}
  \caption{(Color online) Amplitude for Jacobi elliptic  function $u(\zeta)={\rm am}(\zeta |2)$ as solution of the dissipative pendulum equation (\ref{sg-3}) for $c_3=0$.} 
  \label{S-fig3}
 \end{figure}


\begin{figure}[x]
  \centering
  \includegraphics[width= 6.08 cm, height=6.07 cm]{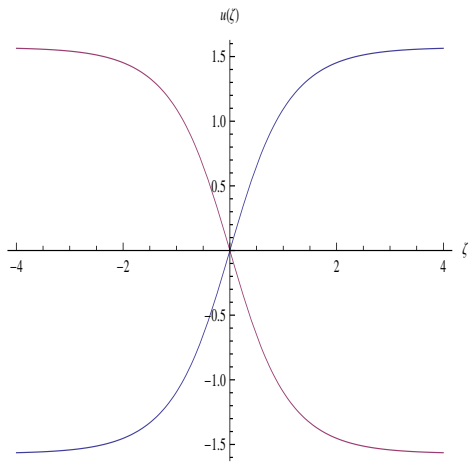}
  \caption{(Color online) Solution $u(\zeta)=\pm {\rm arcsin}\big(\tanh(\sqrt {2}\zeta)\big)$ of the dissipative pendulum equation (\ref{sg-3}) for $c_3=4$.} 
  \label{S-fig4}
 \end{figure}

\begin{figure}[x]
  \centering
  \includegraphics[width= 6.08 cm, height=6.07 cm]{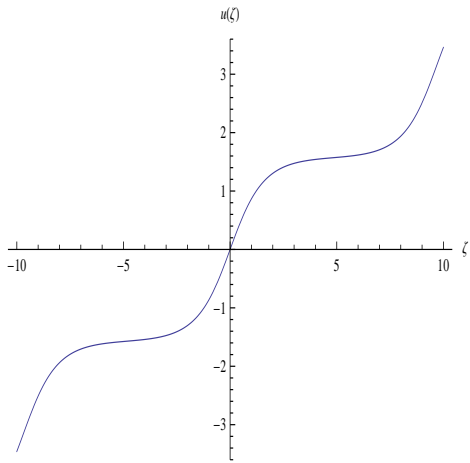}
  \caption{(Color online)  Solution $u(\zeta)={\rm am}(\frac{3 \zeta}{2}|\frac{8}{9})$ of the dissipative pendulum equation (\ref{sg-3}) for $c_3=5$.} 
  \label{S-fig5}
 \end{figure}

\begin{figure}[x]
  \centering
  \includegraphics[width= 6.08 cm, height=6.07 cm]{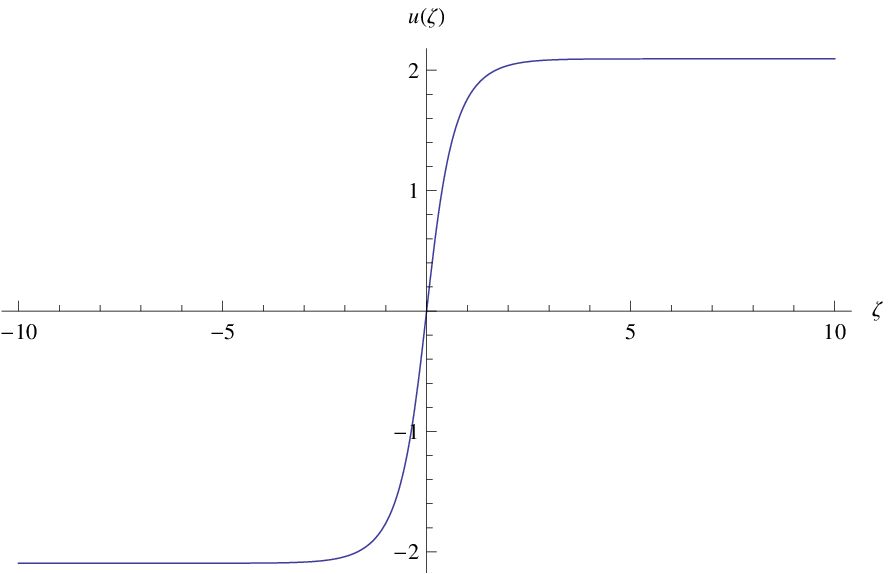}
  \caption{(Color online)  Solution $u(\zeta)$ of the generalized nonlinear pendulum equation (\ref{sine-diss}) for $c_0=1$.} 
  \label{S-fig6}
 \end{figure}

\begin{figure}[x]
  \centering
  \includegraphics[width= 6.08 cm, height=6.07 cm]{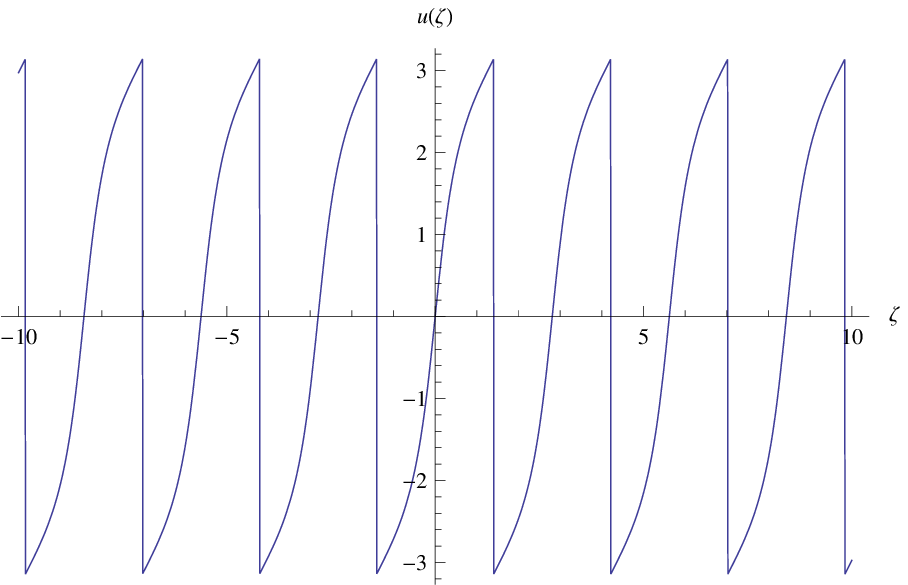}
  \caption{(Color online) Solution $u(\zeta)$ of the generalized nonlinear pendulum equation (\ref{sine-diss}) for $c_0=3$.} 
  \label{S-fig6}
 \end{figure}


\end{document}